\documentclass[aps,pra,nofootinbib,preprint,amsmath,amssymb,floatfix]{revtex4}
\usepackage{color}
\usepackage{graphicx}

\begin{document}

\newcommand{\tc}{\textcolor}
\newcommand{\g}{blue}
\newcommand{\ve}{\varepsilon}
\title{Viscous fluid holographic bounce}         

\author{  I. Brevik$^1$  }      
\affiliation{$^1$Department of Energy and Process Engineering,  Norwegian University of Science and Technology, N-7491 Trondheim, Norway}
\author{A. V. Timoshkin$^{2,3}$}
\affiliation{$^2$Tomsk State Pedagogical University, Kievskaja Street, 60, 634061 Tomsk, Russia}
\affiliation{$^3$Tomsk State University of Control Systems and Radio Electronics, Lenin Avenue, 36, 634050 Tomsk, Russia}

\date{\today}          

\begin{abstract}
We investigate bounce cosmological models in the presence of a viscous fluid, making use of generalized holographic cutoffs  introduced by Nojiri and Odintsov (2017). We consider both an exponential, a power-law, and a double exponential form for the scale factor. By use of these models we calculate expressions for infrared cutoffs analytically, such that they correspond to the particle horizon at the bounce. Finally we derive the energy conservation equation, from the holographic point of view. In that way the relationship between the viscous fluid bounce and the holographic bounce is demonstrated.

\end{abstract}
\maketitle

\bigskip
\section{Introduction}

One interesting way to describe the early universe is to apply the holographic principle. Until recently not may researchers were working in that direction, but a recent study of that kind is the holographic inflationary study of Nojiri {\it et al.} \cite{1}. The rationale for a holographic description of inflation is that, since distances are so small, the holographic energy density becomes necessarily large. The holographic scenario has quite widely been applied in late-time cosmology, in the dark-energy epoch \cite{2,37,39,3} when explaining the acceleration of the universe. In the late-time application
for an accelerating universe there are several choices for the infrared cut-off:  the particle and future horizons, the age of the universe, the inverse square root of the Ricci curvature, or the a combination of Ricci and Gauss-Bonnet invariants. A general infrared cut-off may be constructed as an arbitrary combination all above quantities and their derivatives. Furthermore, if the life-time of the universe is finite due to a Big Rip singularity, then thr infrared radius depends on the  singularity time. The late universe case includes both the matter and holographic dark energy sectors, while in the case of the early universe; the matter sector can be neglected. The theory of holographic dark energy \cite{4,5,6,7,8,9,10,11,12,13} is an effective tool when describing the late-time universe and it turns out to be in good agreements with observations \cite{14,15,16,17,18}.

When turning to early-time cosmology, the holographic principle is applicable with respect to  all the above choices driving the universe evolution. Several investigators have studied the matter bounce model at early times  \cite{19,20,21}. In this picture the universe goes from an accelerated collapse to an expanding era via a bounce, without displaying a mathematical singularity. In this way the universe may experience  a cyclic behavior. After the bounce, the universe enters into a matter-dominated expansion.  General $F(R)$ and $F(T)$ models of gravity in the presence of bounce cosmology were discussed in \cite{22,23,add1,add2}.  The application of holographic theory in connection with a bounce model is of physical interest, as the nonsingular behavior solves in some sense the cosmological singularity problem and becomes an alternative to inflation theory \cite{24,25,26,27,28,29,30}. Bounce cosmology, via an inhomogeneous fluid model, was considered in \cite{31}, and a bounce picture arising from application of the holographic principle in the early universe was investigated in \cite{32}. The current cut-off model of the bounce is a consequence of a generalized future cut-off model.

In the present paper we will apply the holographic principle to the early-time bounce epoch. We will start from the basic  model for the bounce proposed in \cite{22}, and investigate the evolution of the cosmic fluid assuming that the fluid possesses a bulk viscosity. Considering various viscous fluid models we associate infrared cutoffs with particle horizons, and construct the corresponding forms of the energy conservation equations. In that way we establish an equivalence between our viscous fluid bounce model and the holographic bounce model, the latter containing the specific cutoff choice of Nojiri and Odintsov \cite{33}.

One may impose the following question:  does the present theory make any statement about bounce singularities in the future? Does the generalized holographic energy appear in the same way if the universe undergoes successive bounces? We do not think the present theory makes a definite statement about this point, although  in its simplest idealized case one may foresee a cyclic behaviour as indicated above. In reality, there is one important modifying factor that comes from the presence of {\it viscosity}. This means that there is inevitably a loss of kinetic fluid energy, electromagnetic energy, and other kinds of field energy, into heat. The entropy of the universe  has to increase. In this way, the energy gradually has to die out. The second law of thermodynamics makes a simple repetition of bounce events in identical  form impossible.

In the next section we establish the main ingredients in the holographic principle used when describing the universe. Section III discusses cosmological models in the presence of a bounce, showing the mentioned equivalence between viscous-fluid bounce cosmology and holographic bounce. A summary is given in Section IV.

\section{The holographic principle}

We will briefly recall the main content of the holographic principle, following the terminology given, for instance, by Li \cite{15} (it differs from the terminology used in string theory \cite{34,35,36}). An important ingredient is the cutoff radius of the horizon.

Based upon the general holographic energy model proposed by Nojiri and Odintsov \cite{33}, the holographic energy density is taken to be inversely proportional to the squared infrared cutoff $L_{IR}$,
\begin{equation}
\rho=\frac{3c^2}{k^2L_{IR}^2}, \label{1}
\end{equation}
where $k^2=8\pi G$ is Einstein's gravitational constant and $c$ is a parameter (dimensionless in geometric units).

We will investigate the holographic bounce in a homogeneous and isotropic Friedmann-Robertseon-Walker (FRW) metric,
\begin{equation}
ds^2=-dt^2+a^2(t) \sum_{i=1,2,3}(dx^i)^2, \label{2}
\end{equation}

with $a(t)$ the scale factor.

In order to apply the holographic principle to describe bounce cosmology, we may identify the infrared radius $L_{IR}$ with the particle horizon $L_p$, or alternatively, with the future event horizon $L_f$ \cite{37}. These are defined as
\begin{equation}
L_p=a\int_0^t \frac{dt}{a}, \quad L_f= a\int_t^\infty \frac{dt}{a}. \label{3}
\end{equation}

We note that in general, the infrared cutoff $L_{IR}$ could be a function of a combination of both $L_p, L_f$ and their derivatives \cite{13,38} or of the Hubble horizon, the scale factor, and its derivatives \cite{39}.

In the case of early times the first Friedmann equation can be written as
\begin{equation}
H^2=\frac{k^2}{3}\rho, \label{4}
\end{equation}
where $\rho$ is the holographic energy density. The source of this fluid can be imaged to be a scalar field, or it can be related to modified gravity. If we identify $\rho$ in (\ref{4}) with $\rho$ in (\ref{1}), we obtain
\begin{equation}
H=\frac{c}{L_{IR}}.
\end{equation}
We will in this paper assume that the viscous fluid driving the bounce has a holographic origin.

\section{Holographic bounce cosmology via a viscous fluid}

In this section we will consider bounce cosmological models where the scale factor is described by an exponential, a power-law, or a double exponential. Assuming that the bounce has a holographic origin
 we will represent the energy conservation laws in terms of the cutoff radius which, as mentioned above,  is identified with the particle horizon. The motivation for describing the holographic bounce by a viscous fluid, is that the inclusion of viscosity allows one to achieve better agreement between theoretical models and astronomical observations \cite{40,41,42}.

 Let us consider the following form for the equation of state (EoS) of an inhomogeneous viscous fluid in flat FRW spacetime \cite{43,add3,add4,44},
 \begin{equation}
 p=\omega(\rho,t)\rho -3H \zeta (H,t). \label{6}
 \end{equation}
 Here $\zeta(H,t) $ is the bulk viscosity, taken in general to depend on both $H$ and $t$. According to standard thermodynamics, we assume that $\zeta(H,t) >0$.


We will take the following form for the thermodynamic parameter $\omega$ \cite{44},
\begin{equation}
\omega(\rho,t)=\omega_1(t)(A_0\rho^{\alpha-1}-1), \label{7}
\end{equation}
where $A_0>0$ and $\alpha \geq 1$ are constants. Moreover, we choose the bulk viscosity as \cite{45}
\begin{equation}
\zeta(H,t)=\zeta_1(t)(3H)^n, \label{8}
\end{equation}
with $n\geq 0$. It may here be of interest to compare with the following form for $\zeta$ investigated in \cite{normann2017} (with present notation),
\begin{equation}
 \zeta(\rho)=\zeta_0\left( \frac{H}{H_0}\right)^n. \label{8a}
 \end{equation}
where $\zeta_0$ and $\rho_0$ refer to present time. Comparison with observations in \cite{normann2017} favored the choice $n=1$ for the exponent. We see that (\ref{8a}) is simpler than (\ref{8}) in that it corresponds to $\zeta_1=~$constant.

We assume that the universe consists of this one-component viscous fluid, and consider the  energy conservation equation in the form $\zeta(H,t)=3\tau H$,
\begin{equation}
\dot{\rho}+3H(\rho+p)=0. \label{9}
\end{equation}
We will in the following apply the holographic principle to the cosmological models listed above, for different cases of $\omega(\rho,t)$, and for different bulk viscosities $\zeta(H,t)$. In that way we will present the corresponding representations of the energy conservation equation in the holographic picture.

\subsection{Exponential model}

Let us consider a bounce cosmological model where the scale factor has the form
\begin{equation}
a(t)=\exp(\alpha t^2), \label{10}
\end{equation}
with $\alpha$ a positive constant. The instant $t=0$ ($ (\alpha=1$) is when the bounce occurs.

We will start from the simplest case when the EoS parameter is constant, $\omega(\rho,t)=\omega_0$, and when the bulk viscosity is proportional to $H$, $\zeta(H,t)=3\tau H$, with the positive constant $\tau$ having dimension cm$^{-2}$ in geometric units. In this case the EoS equation (\ref{6}) becomes
\begin{equation}
p=\frac{3}{k^2}(\omega_0-3\tau k^2)H^2. \label{11}
\end{equation}
We will describe the holographic bounce in terms of the particle horizon $L_p$   \cite{33}. A calculation of $L_p$ leads to the result
\begin{equation}
L_p=a\int_0^t \frac{dt}{a}=\frac{1}{2}\sqrt{\frac{\pi}{\alpha}}\exp(\alpha t^2){\rm erf} (\sqrt{\alpha}\, t), \label{12}
\end{equation}
where erf$(\sqrt{\alpha}\, t)$ is the probability integral.
For the Hubble function $H$ in terms of the particle horizon $L_p$, and its time derivative, one has \cite{33}
\begin{equation}
H= \frac{\dot{L}_p-1}{L_p}, \quad \dot{H}=\frac{\ddot{L}_p}{L_p}-\frac{\dot{L}_p^2}{L_p^2}+\frac{\dot{L}_p}{L_p^2}. \label{13}
\end{equation}
Then, by using (\ref{11}) and (\ref{13}), the energy conservation law (\ref{9}) in the holographic language takes the form
\begin{equation}
2\left(  \frac{\ddot{L}_p}{L_p}-\frac{\dot{L}_p^2}{L_p^2}+\frac{\dot{L}_p}{L_p^2} \right)+3(\omega_0-3\tau k^2+1)\left(\frac{\dot{L}_p-1}{L_p}\right)^2=0. \label{14}
\end{equation}
Thus, we have presented a simple viscous bounce model based upon the assumption (\ref{10}) for the scale factor.

\subsection{Power-law model}

Next, we consider a cosmological model in which the scale factor varies according to a power-law \cite{22}:
\begin{equation}
a= \bar{a}\left( \frac{t}{\bar{t}}  \right)^q+1, \label{15}
\end{equation}
where $\bar{a}> 0$ is a constant, $\bar{t}$ is a reference time, and $q=2n, \, n \in Z $.

Let us now assume the case of a constant bulk viscosity, $\zeta(H,t)=\zeta_0 >0$. Also, we assume that the thermodynamic parameter $\omega (\rho)$ is a linear function of the energy density,
\begin{equation}
 \omega(\rho)=A_0\rho-1. \label{16}
\end{equation}
The the EoS for the viscous fluid takes the form
\begin{equation}
p=\frac{3}{k^2}H^2\left( \frac{3A_0}{k^2}H^2-1\right)+3\zeta_0 H.
\end{equation}
Again describing the holographic bounce in terms of the particle horizon $L_p$, we obtain
\begin{equation}
\tilde{L}_p^{2n}=-\frac{1}{2n}\sum_{k=0}^{n-1}\cos \left(\frac{2k+1}{2n}\pi\right) \ln \left[\left(\frac{ {\bar a}^{\frac{1}{2n}}t}{\bar t}\right)^2- 2\frac{{\bar a}^{\frac{1}{2n}}t}{\bar t}\cos \left( \frac{2k+1}{2n}\pi\right)+1\right] \nonumber
\end{equation}
\[ +\frac{1}{n}\sum_{k=0}^{n-1}\sin \left( \frac{2k+1}{2n}\pi\right) \arctan \left[ \frac{ {\bar a}^{\frac{t}{2n}}\frac{t}{\bar  t}-\cos \left( \frac{2k+1}{2n}\pi\right)}{\sin \left( \frac{2k+1}{2n}\pi\right)}\right]  \]
\begin{equation}
 +\frac{1}{n}\sum_{k=0}^{n-1}\sin \left( \frac{2k+1}{2n}\pi\right) \arctan \left[\cot \left( \frac{2k+1}{2n}\pi\right)\right], \label{18}
  \end{equation}
  where the relation between $L_p^{2n}$
   and ${\tilde{L}}_p^{2n}$ is
   \begin{equation}
   L_p^{2n} = \frac{\bar{t}}{\bar{a}^\frac{1}{2n}}\left[ \bar{a} \left(\frac{t}{\bar{t}}\right)^{2n}+1 \right]{\tilde{L}}_p^{2n}, \quad n \in N.
   \end{equation}
   In particular, if $n=1$ the expression for $p$  simplifies to
   \begin{equation}
   L_p^2=\frac{\bar{t}}{\sqrt{\bar{a}}}\left[\bar{a}\left(\frac{t}{\bar{t}}\right)^2+1   \right] \arctan{\sqrt{\bar{a}}}\left( \frac{t}{\bar{t}}\right).
   \end{equation}
Then, using (\ref{13}) and  the expression for $p$,   the energy conservation equation (\ref{9}) can be rewritten as
\begin{equation}
2\left(  \frac{\ddot{L}_p}{L_p}-\frac{\dot{L}_p^2}{L_p^2}+\frac{\dot{L}_p}{L_p^2} \right)+3\left( \frac{\dot{L}_p-1}{L_p}\right)^2\left[ \frac{3A_0}{k^2}\left( \frac{\dot{L}_p-1}{L_p}\right)^2 +\frac{\zeta_0k^2L_p}{\dot{L}_p-1} \right] =0. \label{19}
\end{equation}
This is thus the power-law realization of the holographic principle  in the early universe, for the viscous fluid.

\subsection{Double exponential model}

Now we will consider a double exponential model where the scale factor has the form
\begin{equation}
a=\exp(Y)+\exp(Y^2), \label{20}
\end{equation}
where $Y=(t/\bar{t})^2$ and $\bar t$, as before, being a reference time.

We will maintain the linear dependence of the viscosity on the Hubble function, and we will choose a constant $\omega_1(t)=\omega_0$ in the expression (\ref{7}). The EoS then takes the form
\begin{equation}
p=\frac{3}{k^2}H^2\left[ \omega_0\left( \frac{3A_0}{k^2}H^2-1\right) -3\tau k^2 \right]. \label{21}
\end{equation}
A calculation of the particle horizon gives the following result,
\begin{equation}
L_p=\frac{a\bar{t}}{3\sin \alpha}
 \left[ \sin \frac{\alpha}{2}\ln \frac{Y+8^{1/4}\cos \frac{\alpha}{2}\sqrt{Y}+\frac{\sqrt{2}}{2}}
  {Y-8^{1/4} \cos \frac{\alpha}{2}\sqrt{Y}+\frac{\sqrt{2}}{2} }
   +2\cos \frac{\alpha}{2}\arctan \frac{Y-\frac{\sqrt{2}}{2}}
 {\sqrt{2}\sin \frac{\alpha}{2}\sqrt{Y}} \right]
 \end{equation}
 Using the expressions (\ref{13}) and (\ref{21}) the energy conservation law can be written as
 \begin{equation}
2\left(  \frac{\ddot{L}_p}{L_p}-\frac{\dot{L}_p^2}{L_p^2}+\frac{\dot{L}_p}{L_p^2} \right)+3\left( \frac{\dot{L}_p-1}{L_p}\right)^2
\left\{\omega_0 \left[ \frac{3A_0}{k^2}\left( \frac{\dot{L}_p-1}{L_p}\right)^2  -1\right] -3\tau k^2+1  \right\} =0.
 \end{equation}
 This is thus the double exponential model result, as following from a viscous holographic model.

 \section{Conclusion}

 In this paper we  applied the holographic principle to bounce cosmology. The holographic bounce is of phenomenological interest, and has attracted considerable attention. We investigated cosmological models with an exponential, a power-law, or a double exponential form for the scale factor. For each model we calculated analytcally the infrared cutoff, taking it to be equal to the particle horizon. Assuming a bulk viscosity for the cosmic fluid, we calculated the energy conservation equation in each case in the holographic language.

 Thus, we investigated holographic bounce cosmology in the presence of viscosity, and showed the equivalence with the specific cutoff introduced by Nojiri and Odintsov \cite{33}.

\end{document}